\documentclass[aip,rsi,reprint,amsmath,amssymb,a4paper,floatfix]{revtex4-2}
\def\pdftitle{Propulsion Performance Comparison of Hard- and Soft-Magnetic Microrobots Under Rotating Magnetic Fields}
\def\authorname{Joost Wijnmaalen}
\def\pdfsubject{}
\def\pdfkeywords{}
\def\pdfbackref{none}

\usepackage[pdftex,
        colorlinks=true,
        urlcolor=darkblue,               
        anchorcolor=darkblue,
        filecolor=green,                   
        linkcolor=darkblue,              
        menucolor=darkblue,
        citecolor=darkblue,
        pagebackref,
        backref={\pdfbackref},
        pdfpagemode={UseOutlines},
        bookmarks=true,
        bookmarksopen=true,
        pdftitle={\pdftitle},
        pdfauthor={\authorname}, 
        pdfsubject={\pdfsubject},
        pdfkeywords={\pdfkeywords}
        ]{hyperref}

\usepackage{graphicx}
\DeclareGraphicsExtensions{.pdf,.png,.jpg}
\usepackage{thumbpdf}
\usepackage{color}
\definecolor{darkgreen}{rgb}{0,0.5,0}
\definecolor{darkblue}{rgb}{0,0,0.5}
\definecolor{brown}{rgb}{0.98,0.92,0.73}
\definecolor{red}{rgb}{1,0,0}
\definecolor{yellow}{rgb}{1,1,0}
\definecolor{blue}{rgb}{0,0,1}
\definecolor{green}{rgb}{0,1,0}
\definecolor{purple}{rgb}{1,0,1}
\definecolor{gray}{rgb}{0.8,0.8,0.8}
\definecolor{black}{rgb}{0,0,0}
\definecolor{white}{rgb}{1,1,1}
\definecolor{gold}{rgb}{1.,0.84,0.}

\usepackage{amsmath}
\usepackage{wasysym}

\usepackage{booktabs}

\usepackage{pdfsync}

\usepackage{siunitx}

\usepackage{nicefrac}

\usepackage[paper=a4paper,total={170mm,241mm},top=20mm,left=20mm,columnsep=8mm]{geometry}
\usepackage{multirow}
\usepackage{array} 
\usepackage{wrapfig} 

\def\widefigurewidth{\columnwidth}

\newif\ifcmtr
\cmtrfalse
\ifcmtr 
\newcommand{\cmtr}[1]{ %
   [\color{red} \textbf{#1} \normalcolor]%
}%
\else
\newcommand{\cmtr}[1]{ %
}%
\fi

\begin{document}
\title{Comparison of the propulsion of helical microrobots based on
  hard- and soft-magnetic elements under rotating external magnetic fields}

\date{\today}
\author{Joost Wijnmaalen$^{1}$}
\author{Leon Abelmann$^{1}$}
\author{Iulian Apachitei$^{1}$}
\affiliation{ 
  $^1$Delft University of Technology, Delft, The Netherlands.}

\begin{abstract}
\textbf{This study compares the propulsion of helical microrobots
  based on hard- and soft-magnetic elements under rotating magnetic
  fields. Results show that hard-magnetic microrobots achieved
  step-out frequencies and maximum propulsion speeds 4.5 times higher
  than soft-magnetic microrobots. Below saturation magnetization,
  soft-magnetic microrobots demonstrated similar performance
  irrespective of magnetic susceptibility, highlighting that torque
  generation in these materials is purely
  geometry-dependent. Employing a tapered ribbon design increased
  propulsion speed by a factor of 3.5 compared to regular helical
  designs. These results provide a quantitative basis for selecting
  materials and designs, enabling designers to weigh the propulsion
  benefits of hard magnets against the biocompatibility of
  soft-magnetic microrobots.}

\vspace{1em}

\begin{figure}[ht]
    \centering
    \includegraphics[width=\widefigurewidth]{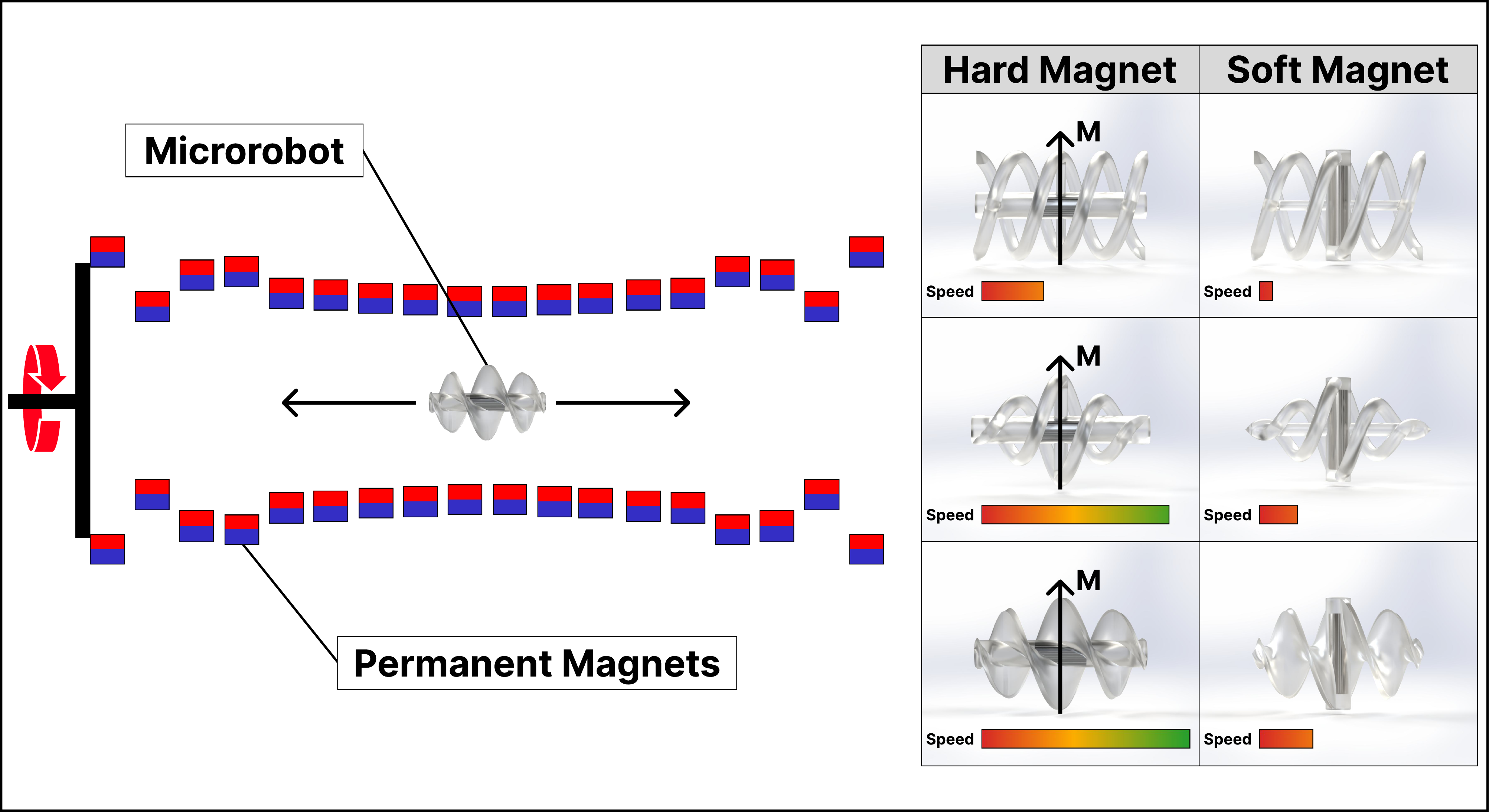}
    \label{fig:Designs}
\end{figure}
\end{abstract}

\maketitle 
\section{Introduction}
Over the past decade, microrobots have emerged as a versatile platform for biomedical applications, including the physical removal of blood clots and biofilms\cite{de_boer_wireless_2025, tran_targeting_2023}, targeted drug delivery\cite{ceylan_3d-printed_2019}, and thermal ablation of bacterial infections\cite{chen_killing_2016}.

Microrobots typically have a polymeric structure created through
two-photon polymerization or micro-stereolithography\cite{li_3d_2021},
and incorporate a magnetic component for torque-driven actuation. This
enables precise microrobot control using tri-axial Helmholtz coil
systems, and also allows for magnetic heating~\cite{nelson_how_2009,
sitti_pros_2020}. The magnetic component can consist of either hard- or
soft-magnetic materials. Hard-magnetic materials, or permanent magnets, 
retain their magnetization over extended periods. They are favored
for their ability to generate high magnetic torques, with NdFeB as a
typical choice~\cite{ren_multi-functional_2019,
hu_small-scale_2018}. However, NdFeB exhibits cytotoxic properties,
necessitating careful consideration for in vivo use and subsequent
removal~\cite{donohue_vitro_1995}. In contrast, soft magnetic
materials lose their magnetic moment quickly. Typical examples like
pure iron or iron-oxide are biocompatible but
provide limited torque output, which might constrain their
use~\cite{ceylan_3d-printed_2019, suter_superparamagnetic_2013}.

Despite these trade-offs, a direct experimental comparison between
hard- and soft-magnetic based microrobots has, to our knowledge, not
been conducted. Such an analysis would contextualize the usability of
magnetic materials already used in microrobots, such as NdFeB and
iron-oxides, and validate the integration of novel materials such as
biodegradable pure Fe, which could be incorporated as thin wires down
to \qty{25}{\micro\meter}. In this study, we compare the
propulsion of hard- and soft-magnetic microrobots under rotating
magnetic fields. 

To enable comparison, three distinct microrobot geometries were each
fabricated with three different magnetic materials: NdFeB
(hard-magnetic), MnZ ferrite (soft-magnetic), and pure Fe
(soft-magnetic). This resulted in nine unique microrobots, all
\qty{10}{\milli\meter} in length. Each was actuated in a
\qty{20}{\milli\tesla} rotating magnetic field. Propulsion tests were
conducted in glycerin, chosen for its high viscosity to achieve low
Reynolds number conditions similar to those expected for smaller
microrobots in future applications~\cite{sitti_biomedical_2015}.


\section{Theory}
\subsection{Magnetic Torque}
The helical microrobots in our study translate rotation around their long axis into forward motion. This rotation is driven by a uniform, externally applied rotating magnetic field. Torque is generated either by embedding a permanent magnet whose magnetization is perpendicular to the direction of motion, or by using a soft-magnetic cylinder oriented with its long axis perpendicular to this direction (see Figure~\ref{fig:Designs}).

\label{sec:MagTorque}
For hard-magnetic microrobots, magnetic torque arises from the tendency of the magnet's fixed magnetization to align with an external magnetic field. Maximum torque is reached when the magnetization and field are perpendicular, and is given by~\cite{furlani_chapter_2001}:

\begin{equation}
  \begin{aligned}
    \label{eq:Tmax_Perm}
    T^\text{hard}_\text{max} = \mu_0 V M H_0 \text{  (\si{\newton\metre}) ,}
  \end{aligned}
\end{equation}

where $\mu_0$ is the vacuum permeability (\si{N/A^2}), $V$ the volume
of the magnetic element (\si{m^3}), $M$ the magnetization (\si{A/m}),
and $H_0$ the applied external magnetic field strength (\si{A/m}).

In soft-magnetic microrobots, torque arises from shape anisotropy. Shape anisotropy depends on geometry, favoring alignment of the material's long axis with the external magnetic field to minimize magnetic energy. When an external magnetic field is applied at an angle to this axis, torque is generated. Maximum torque is reached when the external magnetic field is angled \qty{45}{\degree} relative to the long axis of the soft-magnetic material, and is defined by\cite{abbott_modeling_2007}:

\begin{equation}
  \begin{aligned}
    \label{eq:Tmax_Ferro}
    T^\text{soft}_\text{max} &= \frac{\mu_0 V H_0^2 (n_\text{r} - n_\text{a})}{2 n_\text{a} n_\text{r}} \text{  (\si{\newton\metre}) .}
  \end{aligned}
\end{equation}

Here, $n_r$ and $n_a$ represent the demagnetization factors in the
radial and axial (long) directions, respectively, which can be
approximated with ellipsoid demagnetization factors. This equation is
valid only when the internal magnetization remains below saturation,
defined as:

\begin{equation}
  \begin{aligned}
    \label{eq:Hlow}
    H_\text{low} &= \frac{M_\text{s} n_a n_r \sqrt{2}}{\sqrt{n_a^2 + n_r^2}} \text{  (\si{A/m}) ,}
  \end{aligned}
\end{equation}

where $M_\text{s}$ is the saturation magnetization
(\si{A/m}). By approximating the cylindrical magnets used in this 
study (aspect ratio \num{7.1}) as ellipsoids with an identical 
aspect ratio, saturation fields ($\mu_0 H_\text{low}$) of \qty{23}{mT} 
for ferrite and \qty{86}{mT} for iron were calculated using Osborn's 
equations~\cite{osborn_demagnetizing_1945}. The magnetic field in 
our experiments remains below these values. The torque equations for
$H_0 > H_{\text{low}}$ are provided in Supplementary Material 1
together with the derivations of
Equations~\ref{eq:Tmax_Perm}--\ref{eq:Hlow}.

\subsection{Microrobot Hydrodynamics}
\label{Hydro}

The forward motion of the helical microrobots in viscous fluids (Re
$\ll 1$), is related to their rotation velocity
by~\cite{wang_dynamic_2022}:

\begin{equation}
  \begin{aligned}
    \label{eq:vmax}
    v  = -\frac{b}{a}\omega \text{  (\si{m/s}) .}
  \end{aligned}
\end{equation}

Here, $a$ represents the resistance to forward motion (\si{N.s/m}),
and $b$ denotes the coupling between rotation and translation
(\si{N.s}), both dependent on microrobot geometry and fluid
viscosity. The maximum forward velocity $v_\text{max}$ is limited 
by the highest rotational frequency at which the microrobot can 
maintain synchronous rotation with the external magnetic field, 
known as the step-out frequency $\omega_{\text{max}}$ (\si{Hz}), given by:

\begin{equation}
  \begin{aligned}
    \label{eq:omega_stepout}
    \omega_{\text{max}} = \frac{a}{a c - b^{2}}T_{\text{max}} \text{  (\si{Hz}) ,}
  \end{aligned}
\end{equation}

where $c$ is the rotational resistance (\si{N.s.m}). Above this
rotational frequency, the rotation becomes erratic, and the forward
velocity drops upon further increase of the frequency.

The microrobots in this study were fabricated at a scale one to two
orders of magnitude larger than those envisioned for practical
biomedical applications. Despite this size difference, meaningful
predictions can still be made through extrapolation. This is possible
because, in the laminar flow regime -- characterized by Reynolds
numbers much smaller than unity -- the flow profile is
scale-invariant. To ensure that our millimeter-scale microrobots
operated in this regime, we increased the fluid viscosity by two
orders of magnitude.

To extrapolate our large-scale observations to smaller microrobots, we introduce a scaling factor $\lambda < 1$. When all linear dimensions are scaled by this factor, the overall shape of the microrobot is preserved, and the surrounding flow profile remains geometrically similar.

The drag coefficients $a$, $b$, and $c$ decrease with size and
viscosity as $\lambda\eta$, $\lambda^2\eta$, and $\lambda^3\eta$,
respectively. Similarly, the magnetic torque, whether originating from
a hard (Equation~\ref{eq:Tmax_Perm}) or soft magnet
(Equation~\ref{eq:Tmax_Ferro}), is proportional to the magnet's volume
and thus scales as $\lambda^3$. Substituting these scaling
relationships into Equations~\ref{eq:vmax} and~\ref{eq:omega_stepout}
yields:

\begin{equation}
  \begin{aligned}
    \label{eq:vmax_scaling}
    v_{\text{max}} \propto \frac{\lambda}{\eta} ~\text{.}
  \end{aligned}
\end{equation}

This shows that the maximum speed scales linearly with the size of the
microrobot and inversely with the viscosity of the surrounding fluid. For
example, scaling a \qty{10}{mm} robot down to \qty{100}{\micro\meter}
and reducing the fluid viscosity from \qty{1}{Pa.s} to \qty{5}{mPa.s}
would double \(v_{\text{max}}\) assuming all other parameters remain
unchanged.



\section{Experimental}

\subsection{Rotating Magnetic Field Generation}

\begin{figure}
    \centering
    \includegraphics[width=\widefigurewidth]{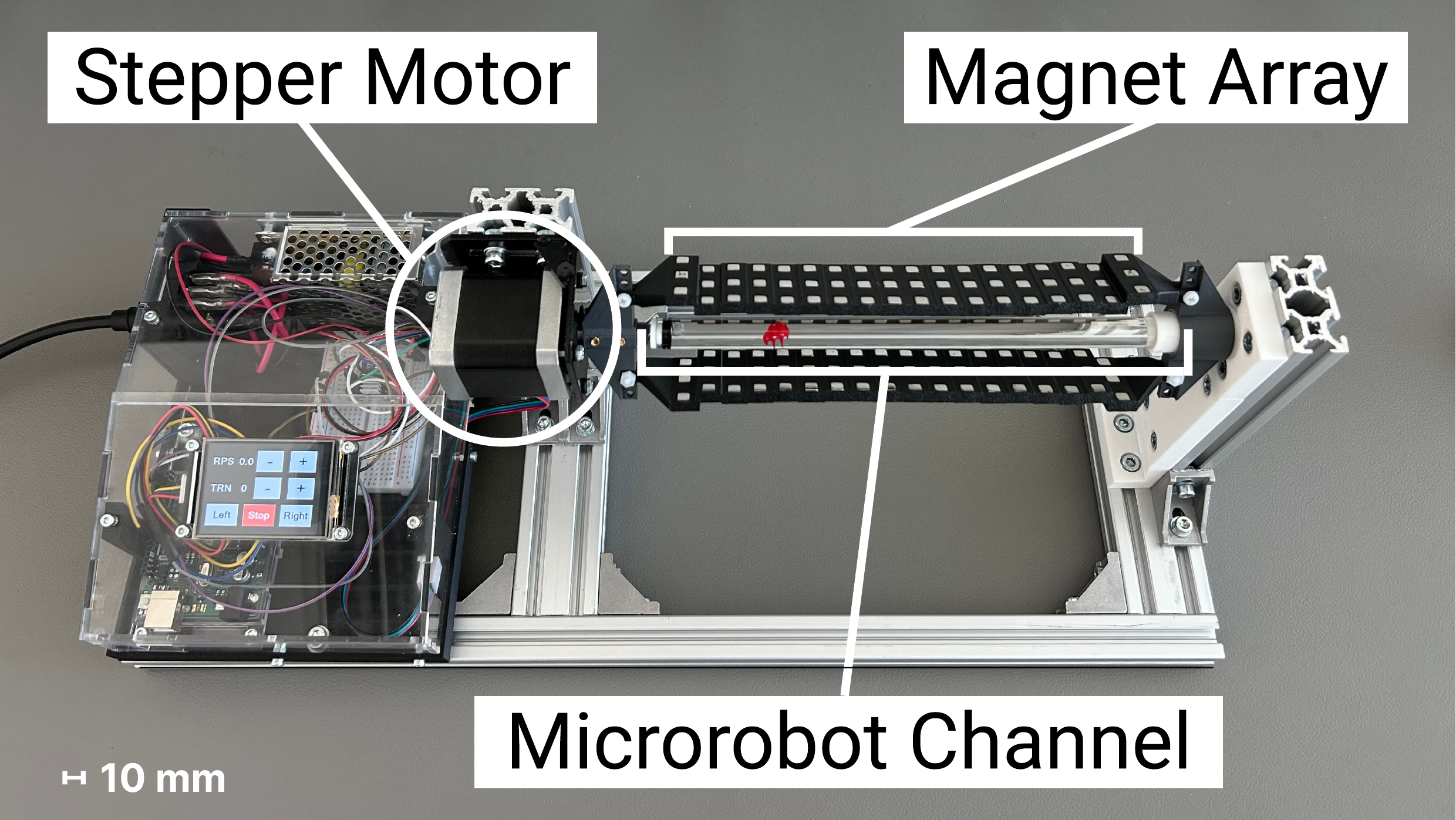}
    \caption{Rotating magnetic field setup for generating uniform
      magnetic torque to propel microrobots. \cmtr{Figures in PDF}}
    \label{fig:Setup}
\end{figure}

The goal of the setup shown in Figure \ref{fig:Setup}, was to generate a uniform rotating magnetic field of \qty{20}{\milli\tesla} over a \qty{120}{\milli\meter} region to enable microrobot propulsion. This system serves as a cost-effective alternative to traditional Helmholtz coil setups. The magnetic field was generated using two identical arrays (175 × \qty{30}{\milli\meter}), each consisting of 3 rows and 18 columns of 5 × 5 × \qty{5}{\milli\meter} N42 NdFeB magnets (Supermagnete, Gottmadingen, Germany). The distance between the opposing magnets was optimized using the Adam algorithm from the Optax library in Python\cite{kingma_adam_2017}. The resulting array was 3D printed using a Bambu Lab X1C, and the individual magnets were secured via press-fit. A uniform field strength of \qty{20}{\milli\tesla} was confirmed (Appendix \ref{app:Field_Validation}) using a Lake Shore 455 DSP Gaussmeter. Additional details about the optimization and construction of the magnetic arrays are included in Supplementary Material 2. 
The magnetic arrays were mounted on a hollow axle which held a channel containing glycerin, with a viscosity of \qty{1.14}{\pascal\second}. The magnetic arrays were rotated using a stepper motor and controlled through a touchscreen connected to an Arduino Uno. To ensure safe operation, the stepper motor was restricted to a maximum rotational frequency of \qty{5}{\hertz}.

\subsection{Microrobot Design and Fabrication}
One hard-magnetic microrobot, NdFeB (N42; first4magnets,
Sutton-in-Ashfield, United Kingdom), and two soft-magnetic
microrobots, MnZ ferrite (78 material; Fair-Rite, Wallkill, United States) and pure Fe (\num{99.5}\% purity; Goodfellow, Huntingdon, United Kingdom), were tested (Figure~\ref{fig:Designs}). Two soft-magnetic microrobots were tested to assess the influence of magnetic susceptibility on torque generation. The NdFeB magnet measured \qty{3}{\milli\metre} in length and \qty{1}{\milli\metre} in diameter and had a remanent magnetization of \qty{1.03}{MA/m}. The ferrite and pure Fe samples were \qty{5.33}{\milli\metre} long and \qty{0.75}{\milli\metre} in diameter to match the magnetic volume of the NdFeB magnet, with magnetic susceptibilities of \num{3000} and \num{8000}, respectively. The saturation magnetization of \num{99.5}\% pure Fe is \qty{1.4}{MA/m} and \qty{0.38}{MA/m} for 78 material ferrite. Three microrobot designs from literature were tested: a double helix, a tapered double helix, and a tapered ribbon-shaped design\cite{lin_magnetic_2021}. All designs were \qty{10}{\milli\metre} in length. The double helix had a constant width of \qty{6.5}{\milli\metre} and a coil diameter of \qty{0.8}{\milli\metre}. The tapered helix had the same length and maximum width but featured a \qty{30}{\degree} taper from both ends toward the center. The ribbon-shaped design consisted of a \qty{0.3}{\milli\metre} thick twisted plate with a central width of \qty{6.5}{\milli\metre}, tapered from both ends toward the center, also at a \qty{30}{\degree} angle.

\begin{figure}
    \centering
    \includegraphics[width=\widefigurewidth]{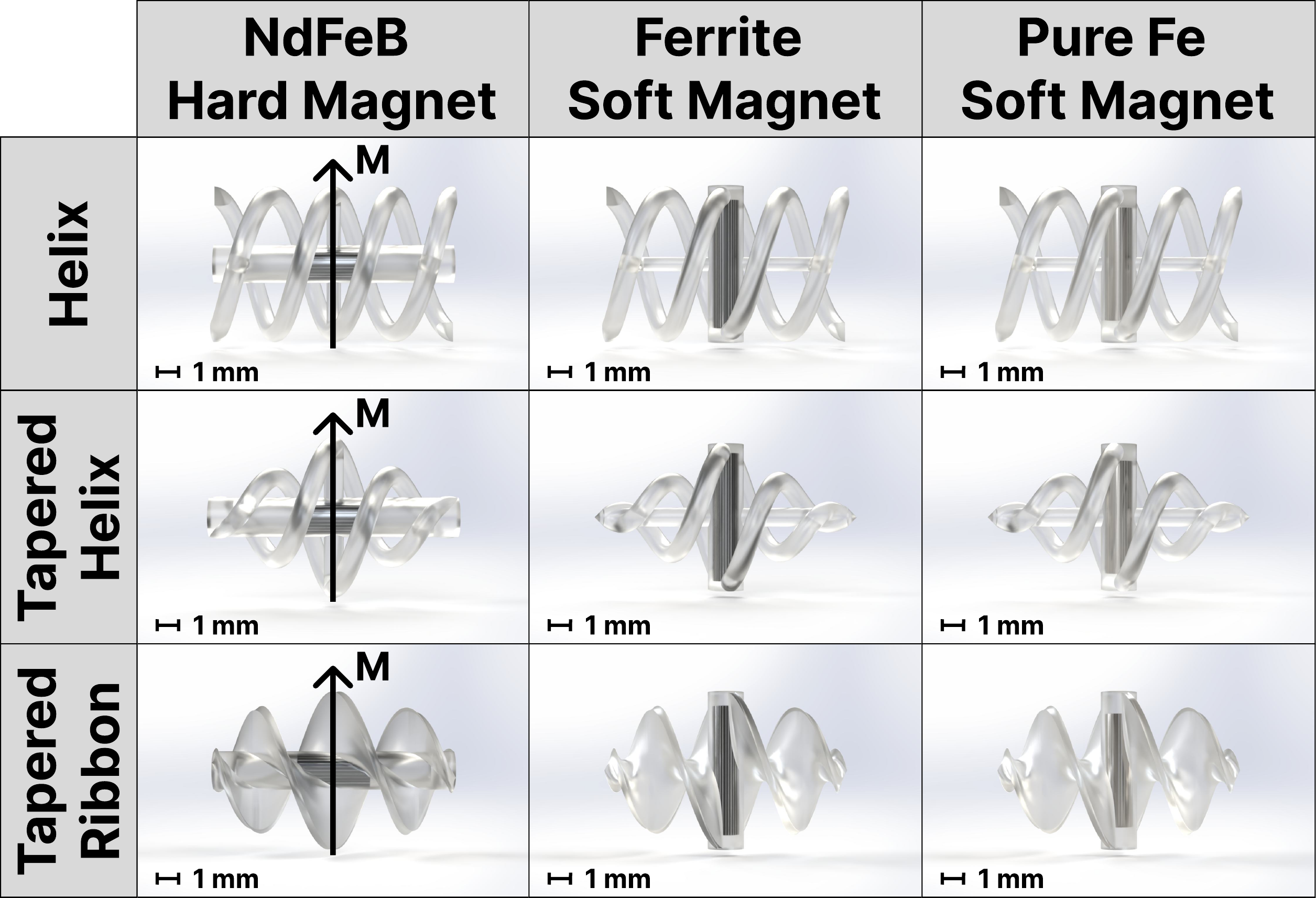}
    \caption{Tested hard- and soft-magnetic microrobot designs, each \qty{10}{\milli\metre} long, with matched magnetic volumes.}
    \label{fig:Designs}
\end{figure}

The microrobots were designed in SolidWorks (Dassault Systèmes, Vélizy-Villacoublay, France), processed using PreForm slicing software (Formlabs Inc., Somerville, MA, USA), and printed on a Formlabs Form 3 stereolithography printer (Formlabs Inc., Somerville, MA, USA) with clear V4 resin. Printed structures were washed in isopropanol (Form Wash, Formlabs; 10 min), post-cured (Form Cure, Formlabs; 30 min), and support marks were manually removed using 800-grit sandpaper. Magnetic materials were secured using UHU Magnet Glue (UHU GmbH \& Co. KG, Bühl, Germany). For full fabrication details, see Supplementary Material 3.

The 3D models of all microrobot designs, corresponding PreForm files, the 3D model of the test setup, wiring diagrams, bill of materials, and optimization code are available on \href{https://github.com/joostwijn/Hard_Versus_Soft_Magnetic_Microrobots}{GitHub}.

\subsection{Microrobot Propulsion Evaluation}
Propulsion characteristics were determined by recording the microrobot’s motion over a \qty{100}{mm} path at \qty{0.1}{\hertz} intervals using an iPhone 14 camera at \qty{60}{fps} (see Supplementary Material 4). The camera was placed at the center of the magnetic arrays at a distance of \qty{120}{mm}. The entry and exit frames were used to determine the travel time and compute the speed. A linear fit was applied to the speed measurements up to the rotation frequency at which half the maximum speed was reached. Subsequently, the \num{95}\% confidence interval of the measurements was determined by:
\begin{equation}
\begin{aligned}
  \label{eq:velocity_uncertainty}
  \Delta v_i &= 1.96\, v_i \sqrt{ \left( \frac{\text{RMSE}}{\bar{v}} \right)^2 + \left( \frac{\sigma_\text{d}}{L} \right)^2 }  ~\text{  (\si{m/s}) ,}
\end{aligned}
\end{equation}
where $\Delta v_i$ is the half-width of the \num{95}\% confidence
interval for the $i$-th speed measurement, $v_i$ is the corresponding
speed, $\text{RMSE}$ is the root-mean-square error of the linear fit
applied to the first \qty{50}{\percent} of data points up to maximum
speed, $\bar{v}$ is the mean speed over that range, $\sigma_\text{d}$
is the distance measurement uncertainty, and $L$ is the nominal
measurement length. The step-out frequency $\omega_\text{max}$ was defined as the lowest rotational frequency at which three consecutive speed measurements, including their confidence intervals, fell entirely below the linear fit.


\section{Results and Discussion}
\subsection{Magnetic Material Comparison}

\begin{table}
  \caption{Comparison of the step-out frequency $\omega_\text{max}$
    and resulting maximum forward speed $v_\text{max}$ with embedded
    hard- or soft-magnetic elements in three different microrobot
    designs. Microrobots based on hard-magnetic elements are 4.5 times
    faster than those based on soft-magnetic elements. The maximum velocity
    for soft-magnetic elements is independent of the material used.}
  \centering
  \label{tab:design_comparison}
  \begin{tabular}{llllllSS}
  \toprule
   \textbf{Design} & \textbf{Magnet} & \textbf{Material} &
    \textbf{$\omega_\text{max}$}& \textbf{$v_\text{max}$} \\
     & \textbf{Type} & & Hz & mm/s\\
    \midrule
    \multirow{3}{*}{Helix} 
                    & Hard      & NdFeB & 4.4(2) & 2.7(2) \\
                    & Soft    & Ferrite & 1.2(1) & 0.6(1) \\
                    & Soft  & Pure Fe & 1.1(1) & 0.6(1) \\
    \midrule
    \multirow{3}{*}{Tapered Helix} 
                    & Hard      & NdFeB & $> 5.0$ & $> 2.8$ \\
                    & Soft    & Ferrite & 3.2(1)  & 1.8(1)  \\
                    & Soft  & Pure Fe & 3.1(1)  & 1.8(1)  \\
    \midrule
    \multirow{3}{*}{Tapered Ribbon} 
                    & Hard      & NdFeB & $> 5.0$ & $> 6.2$ \\
                    & Soft    & Ferrite & 2.4(1)  & 2.1(1)  \\
                    & Soft  & Pure Fe & 2.3(1)  & 2.0(1)  \\
    \bottomrule 
  \end{tabular}%
\end{table}

Table~\ref{tab:design_comparison} summarizes the step-out frequencies and maximum propulsion speeds for all tested materials and designs. Notably, hard-magnetic (NdFeB) microrobots achieved higher step-out frequencies and corresponding maximum speeds compared to soft-magnetic (ferrite and pure Fe) variants. 
Table~\ref{tab:design_comparison} also shows that both soft-magnetic materials exhibited similar propulsion performance despite having different magnetic susceptibilities. This aligns with the findings of Abbott \textit{et al.}\cite{abbott_modeling_2007}, who reported that torque generation in soft-magnetic materials is insensitive to magnetic susceptibility. This holds true as long as the magnetic material remains unsaturated. Since commercially available Helmholtz coil systems typically operate below \qty{20}{\milli\tesla}, magnetic saturation is unlikely to occur when using such systems for microrobot propulsion.

\subsection{Microrobot Design Comparison}
\begin{figure}
    \centering
    \includegraphics[width=\widefigurewidth]{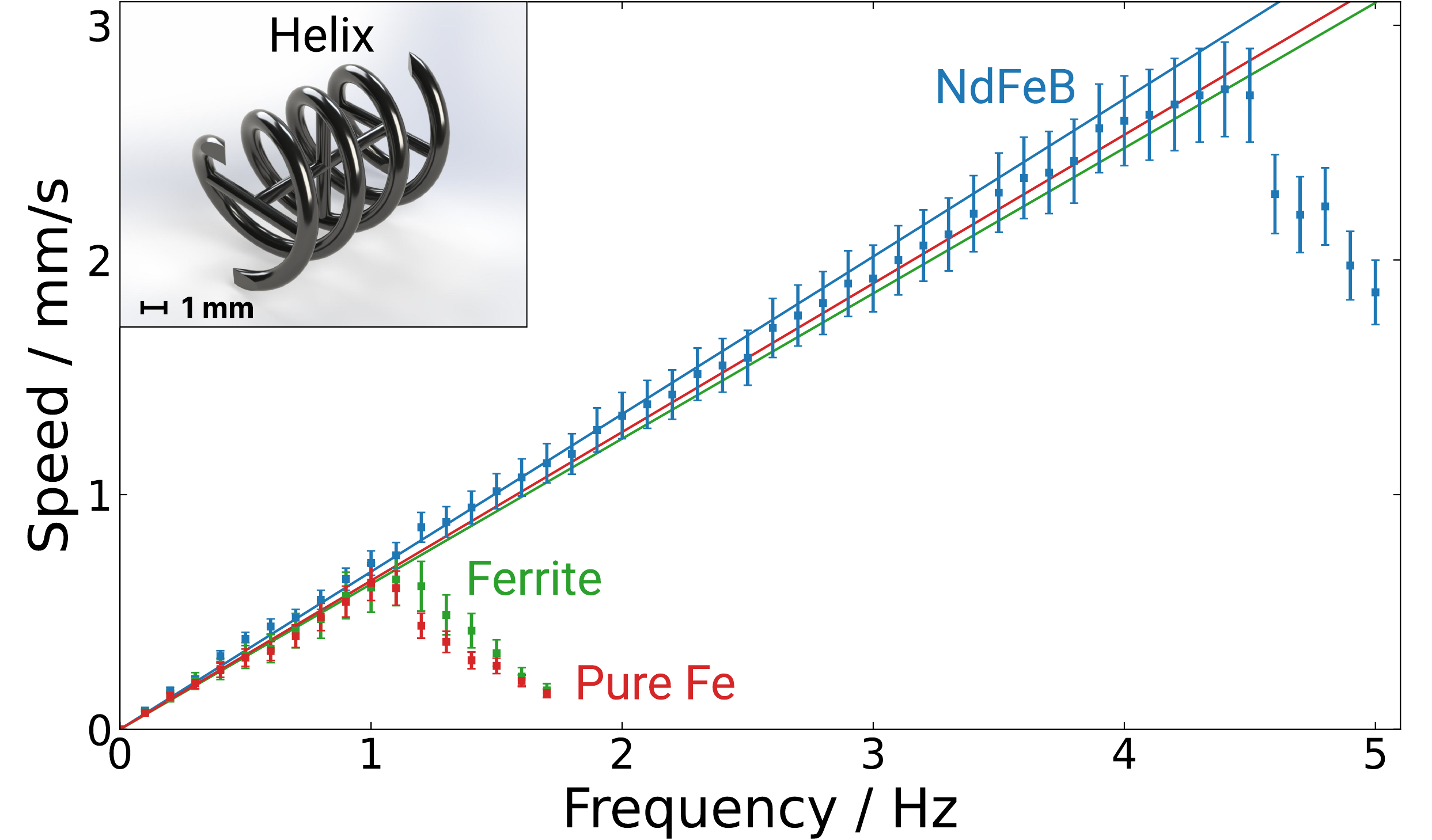}
    \caption{Measured translation speed as a function of external magnetic field rotation frequency for one hard-magnetic (NdFeB) and two soft-magnetic (ferrite and pure Fe) helical microrobots. All designs exhibited similar linear behavior up to their step-out frequency, with the hard-magnetic helix achieving a step-out frequency and maximum speed 4.5 times higher compared to the soft-magnetic designs.}
    \label{fig:Helix_Performance}
\end{figure}

Figure~\ref{fig:Helix_Performance} shows that hard-magnetic helical
microrobots achieved a maximum speed \num{4.5} times higher than both
soft-magnetic designs, which aligns with torque predictions based on
Equations~\ref{eq:Tmax_Perm} and~\ref{eq:Tmax_Ferro} (see
Supplementary Material 1). The nearly identical propulsion efficiencies (slope: 0.64(3)) of the hard- and soft-magnetic helical designs also indicate that the orientation of the magnetic material, whether along the length or width of the microrobot, has minimal effect on propulsion characteristics. Interestingly, the relationship between rotational frequency of the magnetic field and translational speed became nonlinear approaching maximum speed. This behavior contrasts with the fully linear response reported by Wang \textit{et al.}~\cite{wang_dynamic_2022}. The observed gradual loss of linearity likely results from subtle local variations in the magnetic field or from local increases in drag due to occasional wall contact. Both effects can cause the microrobot to reach the step-out frequency earlier in specific regions along its propulsion path. However, these deviations remained small, indicating minimal wall contact and underscoring the overall uniformity of the generated magnetic field.

\begin{figure}
    \centering
    \includegraphics[width=\widefigurewidth]{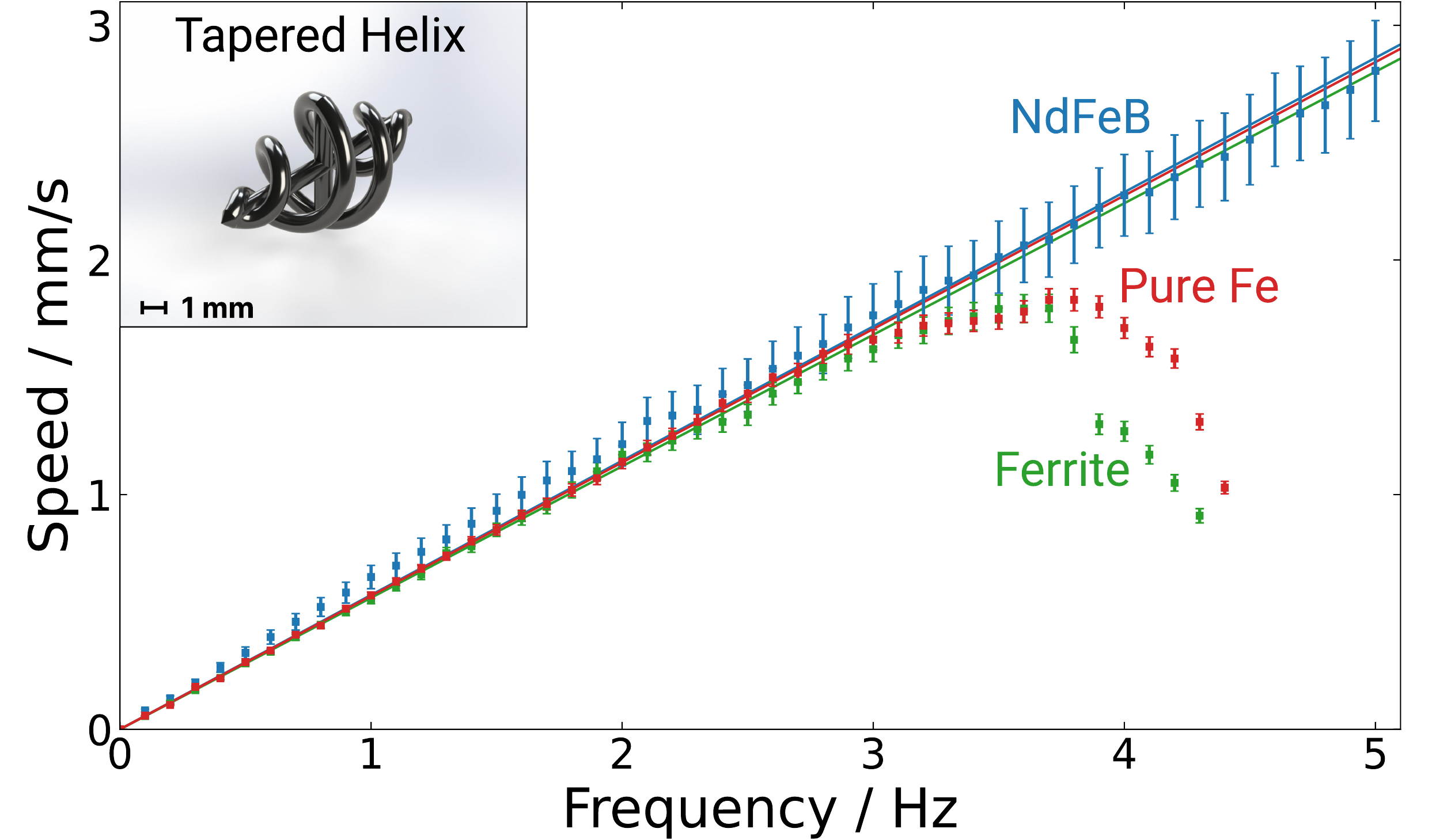}
    \caption{Measured translation speed as a function of external magnetic field rotation frequency for one hard-magnetic (NdFeB) and two soft-magnetic (ferrite and pure Fe) tapered helical microrobots. This design achieved a threefold increase in step-out frequency and maximum speed compared to regular helical designs.}
    \label{fig:Tap_Helix_Performance}
\end{figure}

The tapered helix achieved the highest step-out frequency among all
soft-magnetic designs, but also had the lowest propulsion efficiency
(slope: (0.57(1)), as can be seen in
Figure~\ref{fig:Tap_Helix_Performance}. This enhancement in speed and
step-out frequency is attributed to the reduced width and increased
spacing between the microrobot and surrounding surfaces, which
decreases rotational friction. Although this design slightly
compromises propulsion efficiency, the threefold increase in maximum
speed outweighs this reduction. It should be noted that the step-out
frequency of hard-magnetic tapered helices exceeded the experimental
limit of \qty{5}{Hz}, preventing direct confirmation of the three-fold
increase in step-out frequency and maximum speed. Reducing the magnetic field strength could lower the step-out frequency to within the experimental limit of \qty{5}{Hz}. However, this would also narrow the operational frequency range of the soft-magnetic microrobots, reducing result accuracy. Therefore, this approach was not pursued. Given the
identical slopes and torque-dependent propulsion behavior observed in
the regular helical designs, however, it is likely that a similar
threefold increase applies to the hard-magnetic tapered helical
designs. Furthermore, the higher rotational speeds of this design may
also improve the physical ablation of structures with shear-thinning
properties, such as
biofilms~\cite{prades_computational_2020}. However, further research
is needed to quantify this effect.

\begin{figure}
    \centering
    \includegraphics[width=\widefigurewidth]{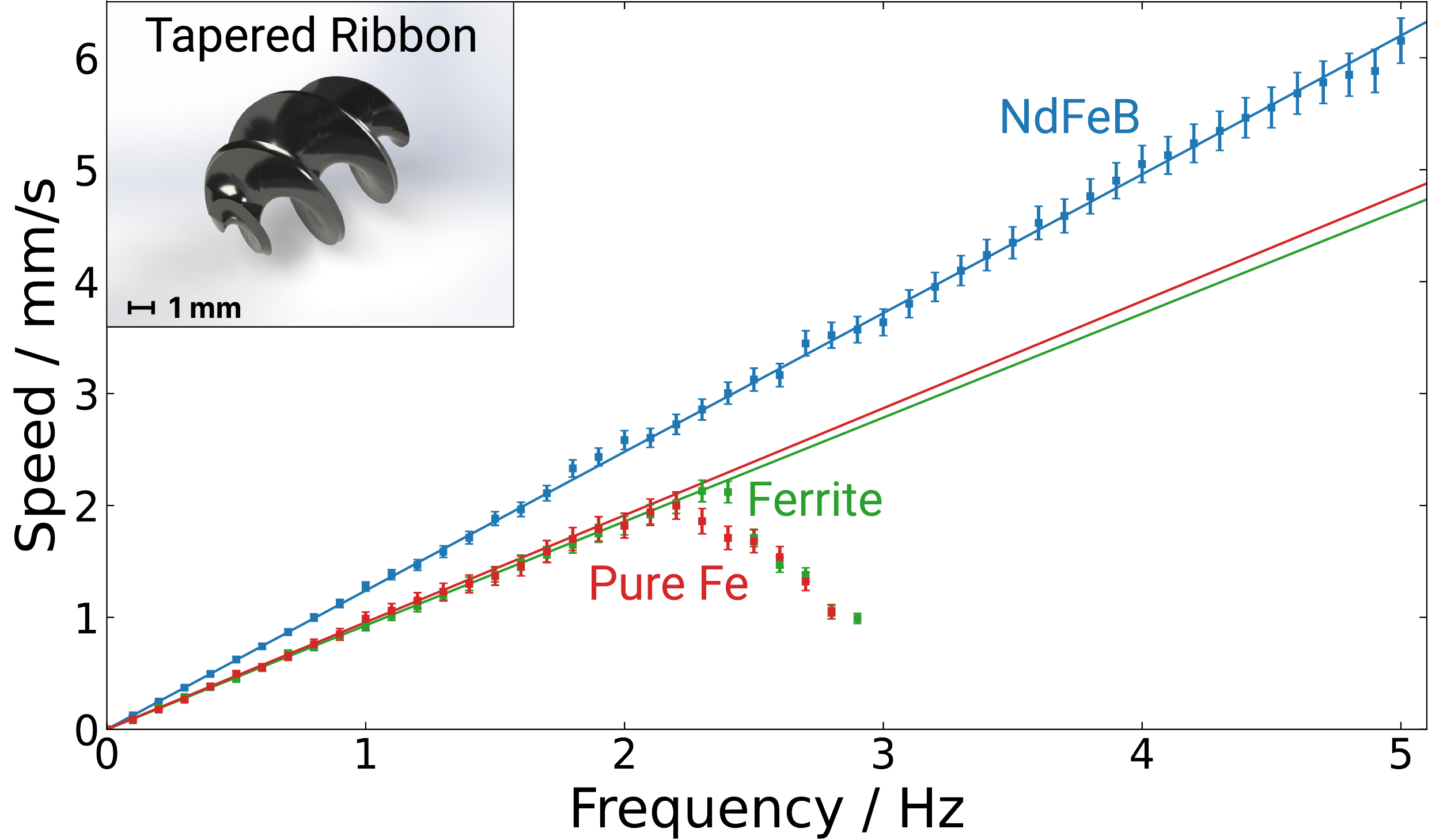}
    \caption{Measured translation speed as a function of external magnetic field rotation frequency for one hard-magnetic (NdFeB) and two soft-magnetic (ferrite and pure Fe) ribbon-shaped microrobots. This design achieved the highest maximum speeds among all designs.}
    \label{fig:Tapered_Ribbon}
\end{figure}

Figure~\ref{fig:Tapered_Ribbon} shows that the tapered ribbon achieved the highest maximum speed among all tested designs, which could shorten procedure times in biomedical applications. Its step-out frequency was twice that of the regular helix, while its maximum speed was \num{3.5} times greater. Despite having a step-out frequency lower than the tapered helix, the tapered ribbon achieved the highest maximum speed. This underscores the intricate relationship between rotational resistance and propulsion. The larger surface area of the ribbon-shaped design may also allow for greater drug loading and faster release. Interestingly, a small but statistically significant efficiency gap was observed between hard- and soft-magnetic ribbon-shaped designs, with slopes of 1.24(2) and 0.95(2) respectively. This likely occurred because the thinner ribbon walls made the transverse rods in the soft-magnetic designs contribute more to drag. In contrast, the thicker overall structure of the helical designs resulted in minimal additional drag from the transverse rod. Nevertheless, the ribbon-shaped designs achieved the highest maximum speed of all tested microrobots.


\section{Conclusions}
We compared the propulsion of helical microrobots actuated by an
external rotating magnetic field, using either hard- or soft-magnetic
elements and three representative shapes. This allowed us to isolate
the effects of magnetic material and geometry on maximum propulsion speed.

Helical microrobots incorporating hard-magnetic materials outperform
their soft-magnetic counterparts under rotating magnetic fields,
achieving a \num{4.5}-fold higher step-out frequency and corresponding
maximum propulsion speed under matched magnetic volume, field
strength, and viscosity at low Reynolds numbers. Notably, for
soft-magnetic materials, the generated torque is independent of
magnetic susceptibility, resulting in identical propulsion
characteristics for comparable geometric designs for external magnetic
field strengths below the saturation value of approximately
\qty{23}{mT} for ferrite and \qty{86}{mT} for pure Fe. Furthermore,
employing a tapered ribbon-shaped design was shown to improve maximum
propulsion speed by a factor of \num{3.5} compared to a regular
helical design, underscoring the influence of geometry on microrobot
performance.

These performance differences provide a quantitative basis for
informed material and design selection. This enables designers to
balance the propulsion advantages of hard magnets against the
biocompatibility of soft-magnetic alternatives. It also allows them to
select the most suitable geometric design for their
application.

This work was limited to propulsion along a line. Future work should
explore navigation in more dimensions, as well as microrobot
functionalities beyond propulsion, with a particular focus on heating
efficiency for thermal ablation and controlled drug release. Comparing
hard- and soft-magnetic materials in this context would offer a more
comprehensive understanding of their respective advantages and
limitations, ultimately guiding the design and application of
microrobots for specific clinical tasks.


\section*{Acknowledgments}
The authors thank Sander Leeflang and Jinlai Li for their assistance with 3D printing the microrobots.

\section*{References}
\bibliographystyle{apsrev4-2-modified}
%

\clearpage
\appendix
\section{Magnetic Field Validation}
\label{app:Field_Validation}
\begin{figure}[h!]     
    \centering
    \includegraphics[width=\widefigurewidth]{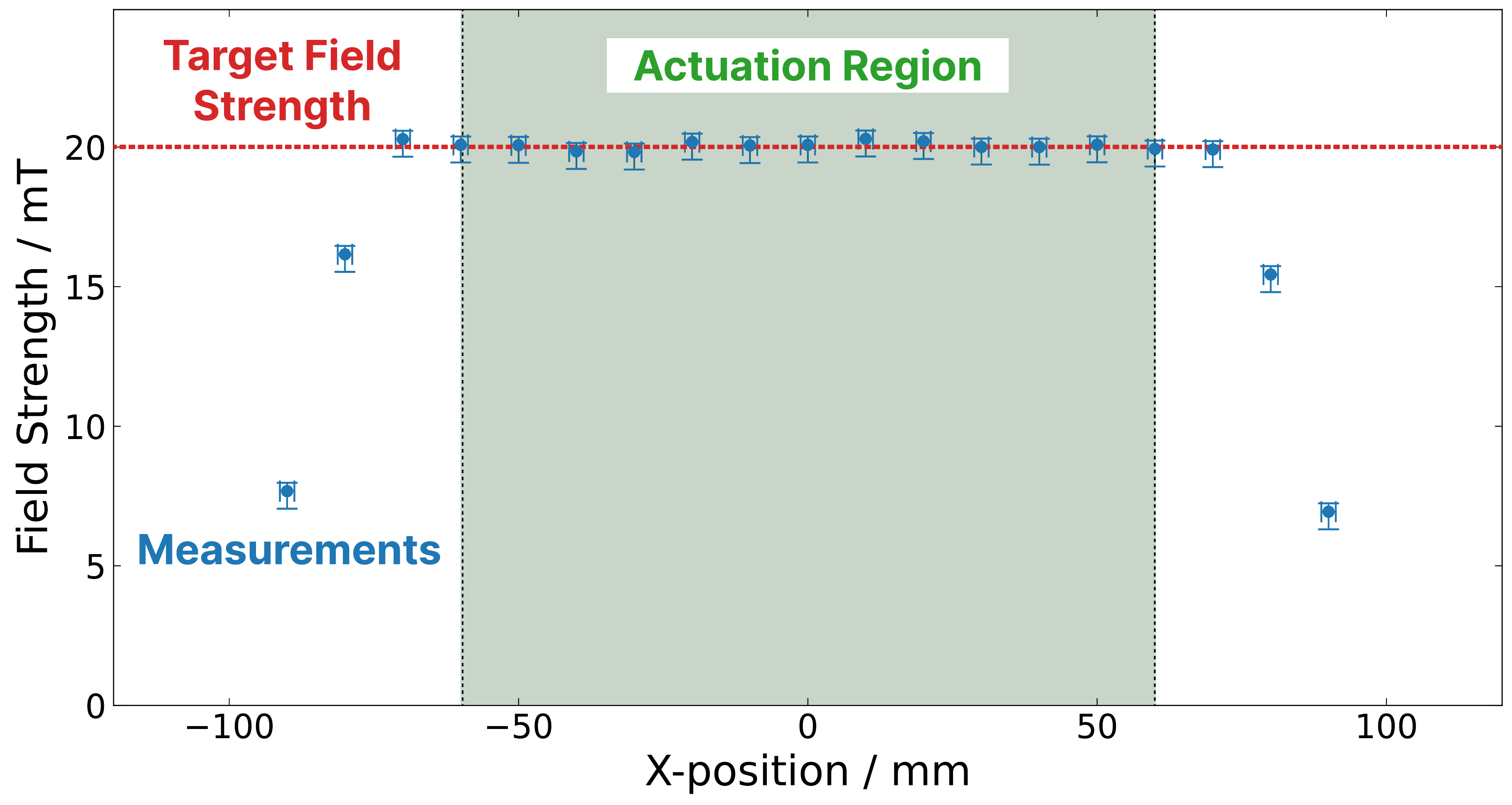}  
    \caption{Measured magnetic field strength along the central axis of the rotating magnetic field setup (x-axis). A \qty{20}{\milli\tesla} field was maintained across the \qty{120}{\milli\metre} wide actuation region, with little deviation from the target field.}
    \label{fig:Magnetic_Field}
\end{figure}

\end{document}